\documentclass[aps, pre, reprint, superscriptaddress, floatfix, longbibliography, nobibnotes]{revtex4-2}

\usepackage[utf8]{inputenc}
\usepackage{amsmath}
\usepackage{amsfonts}
\usepackage{amssymb}
\usepackage{siunitx}
\usepackage{physics}
\usepackage{stmaryrd}
\usepackage[colorlinks=true, citecolor=blue, urlcolor=red]{hyperref}
\usepackage{orcidlink}

\newcommand{\mean}[1]{\langle {#1} \rangle}
\newcommand{\PI}{P_\mathrm{I}}
\newcommand{\PII}{P_\mathrm{II}}
\newcommand{\PIII}{P_\mathrm{III}}

\DeclareMathOperator{\erfc}{erfc}
\newcommand{\tfp}{t_\mathrm{fp}}
\newcommand{\e}{\mathrm{e}}
\renewcommand{\d}{\mathrm{d}}
\newcommand{\trelax}{\tau_r}

\begin{document}
\title{First passage time distribution in underdamped harmonic oscillators}

\author{Aubin Archambault\, \orcidlink{0009-0002-3373-357X}}
\affiliation{\href{https://ror.org/02feahw73}{CNRS}, \href{https://ror.org/04zmssz18}{ENS de Lyon}, \href{https://ror.org/00w5ay796}{Laboratoire de Physique}, F-69342 Lyon, France}
\author{Caroline Crauste-Thibierge\, \orcidlink{0000-0001-5502-0445}}
\affiliation{\href{https://ror.org/02feahw73}{CNRS}, \href{https://ror.org/04zmssz18}{ENS de Lyon}, \href{https://ror.org/00w5ay796}{Laboratoire de Physique}, F-69342 Lyon, France}
\author{Alberto Imparato\, \orcidlink{0000-0002-7053-4732}}
\affiliation{Department of Physics, University of Trieste, Strada Costiera 11, 34151 Trieste, Italy}
\affiliation{ Istituto Nazionale di Fisica Nucleare, Trieste Section, Via Valerio 2, 34127 Trieste, Italy}
\author{Sergio Ciliberto\, \orcidlink{0000-0002-4366-6094}}
\affiliation{\href{https://ror.org/02feahw73}{CNRS}, \href{https://ror.org/04zmssz18}{ENS de Lyon}, \href{https://ror.org/00w5ay796}{Laboratoire de Physique}, F-69342 Lyon, France}
\author{Ludovic Bellon\, \orcidlink{0000-0002-2499-8106}}
\email{ludovic.bellon@ens-lyon.fr}
\affiliation{\href{https://ror.org/02feahw73}{CNRS}, \href{https://ror.org/04zmssz18}{ENS de Lyon}, \href{https://ror.org/00w5ay796}{Laboratoire de Physique}, F-69342 Lyon, France}

\date{\today}

\begin{abstract}
We derive the distribution of the first passage time $\tfp$ for the position $x$ of an underdamped harmonic oscillator to overcome a threshold $x_B$. As the $\tfp$ distribution depends on the oscillator quality factor $Q$ different approaches are used. At very large quality factor ($Q\gg 100$) and intermediate and long $\tfp$ the proof is based on an energy diffusion model, whereas at medium quality factor ($Q\sim 10$) the proof is based on the study of the eigenvalues of the Kramers linear differential operator with absorbing boundary conditions. For all $Q$ and short $\tfp$ we use a Hamiltonian approximation. The theoretical predictions are in excellent agreement with direct numerical simulations of underdamped oscillator dynamics. Finally we show that the mean of the trajectories ending at $\tfp$ presents a particular shape driven by a specific noise pattern.
\end{abstract}

\maketitle
\section{Introduction} \label{Sec_Intro}
The knowledge of the First Passage Time distribution (FPTd)~\cite{Redner_2001} is useful in a large variety of problems ranging from astrophysics~\cite{Chandrasek} to chemistry~\cite{Bray_2013, Hanggi_1990}, from computer science~\cite{Majumdar2005} to physics and biology~\cite{Godec2017, Shin2019}. Previous theoretical~\cite{Majumdar2005, Majumdar_2008, Evans2011, Evans2021} and experimental~\cite{Roichman_2018, Roichman_2020, Roichman_2025, Besga_2020, Besga_2021, Faisant_2021} studies have been developed for overdamped systems and a full knowledge of FPTd lacks for the underdamped ones such as for example harmonic oscillators with a small damping (i.e. large quality factor $Q$). The reason is that in these underdamped systems the position dynamics has long memory effects because of inertia. Thus the calculation of FPTd may become very complex~\cite{Bray_2013} and the results have a strong dependency on $Q$. 

In an accompanying Letter~\cite{FPTuSHO_Letter}, we derive the FPTd of harmonic oscillators in the case of intermediate $Q$. The result is obtained by following different approaches for short and long times that we will recall in the following sections. The theoretically estimated FPTd agrees with the experimental one measured on an oscillator with $Q=7$. The computed mean FPT is used to estimate the power of an underdamped information engine finding an excellent agreement with experimental values. 

In this article we report the details of the calculations and we extend them to very large $Q$, which is a useful result for oscillators with extremely small damping. Furthermore we analyze the mean trajectories leading to the FPT and we show that they correspond to specific noise patterns. All theoretical results are supported by numerical simulation illustrations. The article is organized as follow. In Sec.~\ref{Sec_time_pdf} we describe the system and we recall the main properties of the FPT distribution. In Sec.~\ref{Sec_FPT_formalism} we explain the formalism used to derive the FPTd for large times at any $Q$, based on finding the slowest eigenvalue of the linear operator of Kramers' equation describing the phase space dynamics. In Sec.~\ref{Sec_large_Q} we study the limiting case of large $Q$ and develop an alternative approach to obtain the FPT distribution. In Sec.~\ref{Sec_noise_pattern} we study the recurrent noise pattern that appear just before the FPT using the instanton theory. In Sec.~\ref{Sec_short_FPT} we qualitatively analyze how our framework also accounts for the transient behavior at short FPT. Finally we conclude in Sec.~\ref{Sec_Conclu}.

\section{Trigger time distribution}\label{Sec_time_pdf}

We consider a harmonic oscillator characterized by its mass $m$, stiffness $k$, resonance angular frequency $\omega_0=\sqrt{k/m}$, viscous damping $\gamma$ and quality factor $Q=m\omega_0/\gamma$, in equilibrium with a thermostat at temperature $T$. We define the unit length as $\sigma=\sqrt{k_BT/k}$ (with $k_B$ the Boltzmann constant), the unit time as $\omega_0^{-1}$, and the unit of energy as $k_BT$. The harmonic potential is then $U(x)=\frac{1}{2}x^2$, and the Langevin equation describing the dynamics of the position $x$ and velocity $v=\dot x$ reads:
\begin{subequations}\label{Lan:eq}
\begin{align}
 \dot x &= v, \\
 \dot v &= -\frac{1}{Q} v - U'(x) + \sqrt{\frac{2}{Q}} \eta, 
\end{align}
\end{subequations}
with $\eta$ a delta correlated noise of unit variance. The relaxation time in position or velocity of the harmonic oscillator is $\trelax=2Q$ ($2Q/\omega_0$ in dimension-full units).

\begin{figure}[htb]
	\centering
	\includegraphics[width=\columnwidth]{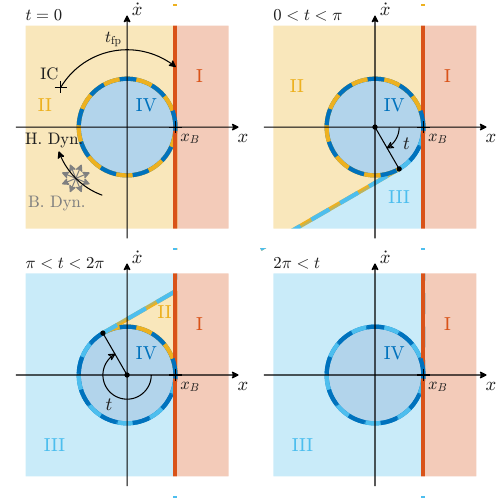}
	\caption{\label{Fig:PhaseSpacexv} Phase space $(x, v)$ of the harmonic oscillator, at four different times. The Hamiltonian dynamics ($Q=\infty$) correspond to a clockwise rotation at constant speed of any initial condition (IC), while thermal noise (Brownian dynamics, when $Q$ is finite) adds a diffusion around this deterministic motion. All initial conditions in the red shaded area (I, $x(t=0)\ge x_B$) overcome the threshold instantaneously and correspond to $\tfp=0$. The blue-yellow straight line indicates the evolution of the initial boundary (red vertical straight line) as a function of the phase space rotation. During the time evolution, all initial conditions in the yellow shaded area (II) cross the threshold at a constant rate as long as the first passage time is $\tfp\le\pi$. Their number decreases when $\pi<\tfp<2\pi$, as the extension towards large energies is bounded by $E^\dagger$ and tends to 0 for $\tfp=2\pi$. Behind region II is the light blue shaded area labeled III, that corresponds solely to initial conditions diffusing from the area IV (dark blue shaded), defined by $E(t=0)<B$. Area IV is only emptied through area III, at Kramers' escape rate, and gradually replace region II till $\tfp\ge2\pi$, when it corresponds to the only first passage time mechanism. This picture is supported by the movies of the phase space dynamics based on numerical simulations available as ancillary files~\cite{SuppMatMovies}.}
\end{figure}

Starting from the equilibrium distribution of position, we are interested in computing the distribution $P(\tfp)$ of the first passage time $\tfp$ for $x$ to overcome a threshold $x_B$. Such event implies that the total energy of the resonator $E=\frac{1}{2}v^2+\frac{1}{2}x^2$ in the harmonic well is larger than $B=\frac{1}{2}x_B^2$. As described in the companion Letter~\cite{FPTuSHO_Letter}, the distribution $P(\tfp)$ is the sum of three contributions:
\begin{equation} \label{Eq:pdftfp}
	P(\tfp) =\PI\delta(\tfp)+\PII(\tfp)+\PIII(\tfp), 
\end{equation}
with
\begin{subequations} \label{Eqs:PI-PII-PIII}
\begin{align}
 \PI &= \frac{1}{2}\erfc\left(\sqrt{B}\right), \label{EqPI}\\
 \PII &= \frac{1}{2\pi}\left(\e^{-B}-\e^{-E^\dagger(\tfp)}\right), \label{EqPII}\\
 \PIII &= (1-\e^{-B})\Gamma(\tfp)\exp\left(-\int_0^{\tfp} \Gamma(t)dt\right), \label{EqPIII}
\end{align}
\end{subequations}
where $\PI$ the instantaneous contribution corresponding to $x(t=0)\ge x_B$ (red shaded area I in Fig.~\ref{Fig:PhaseSpacexv}); $\PII$ the short time contribution for the remaining large initial energies $E(t=0)\ge B$ (yellow shaded area II in Fig.~\ref{Fig:PhaseSpacexv}); and $\PIII$ the long time contribution for the initial conditions $E(t=0)< B$ (blue shaded area IV in Figs.~\ref{Fig:PhaseSpacexv}). In Eq.~\ref{EqPII}, $E^\dagger(\tfp)$ is defined as
\begin{subequations} \label{Eq:Edagger}
	\begin{align} 
		E^\dagger(\tfp)&=\infty &&\text{if } 0<\tfp\le \pi, \label{Eq:Edaggera}\\
		E^\dagger(\tfp)&=\frac{B}{\cos^2(\tfp/2)} &&\text{if } \pi<\tfp\le 2\pi, \label{Eq:Edaggerb}\\
		E^\dagger(\tfp)&=B &&\text{if } 2\pi<\tfp. \label{Eq:Edaggerc}
	\end{align}
\end{subequations}

Finally, in Eq.~\ref{EqPIII}, $\Gamma$ is the Kramers escape rate to cross the energetic barrier $B$ and eventually reach $x_B$. As a rule of thumb, $\Gamma$ can be estimated by considering that the oscillator probes the equilibrium Boltzmann energy distribution and overcomes the barrier $B$ with a probability $P(E>B)\sim\e^{-B}$ each time it forgets about its previous energetic state, thus with a frequency $\sim 1/Q$~\cite{note-tauE}:
\begin{equation}
 \Gamma \sim \Gamma_K \equiv \frac{1}{Q}\e^{-B}.
\end{equation}

We work in the limit of low damping ($Q\gg 1$), so that $E$ can be considered constant during one oscillation period (Hamiltonian approximation). In the phase space $(x, v)$, this correspond to a circular trajectory of radius $x_E=\sqrt{2E}$ at constant unit angular speed. Brownian noise appears as a diffusion around these circular trajectories. The origin of the three contributions to the pdf of first passage time is pictured in Fig.~\ref{Fig:PhaseSpacexv}, which illustrates the short time contributions discussed in Ref.~\onlinecite{FPTuSHO_Letter}. In the following, we essentially focus on the long time limit ($\tfp>2\pi$), corresponding to the bottom right plot: reaching $x_B$ from initial conditions at low energy. This amounts to solving a Kramers escape problem, with the system that in principle can cross the energy barrier $B$ back and forth several times before reaching the target position $x_B$ with some non-negative velocity.

\section{First Passage time formalism}\label{Sec_FPT_formalism}

In this section, we review the first passage time formalism for a Brownian particle with inertia described by the Langevin equation \eqref{Lan:eq}.

\begin{figure*}[t]
 \centering
 \includegraphics{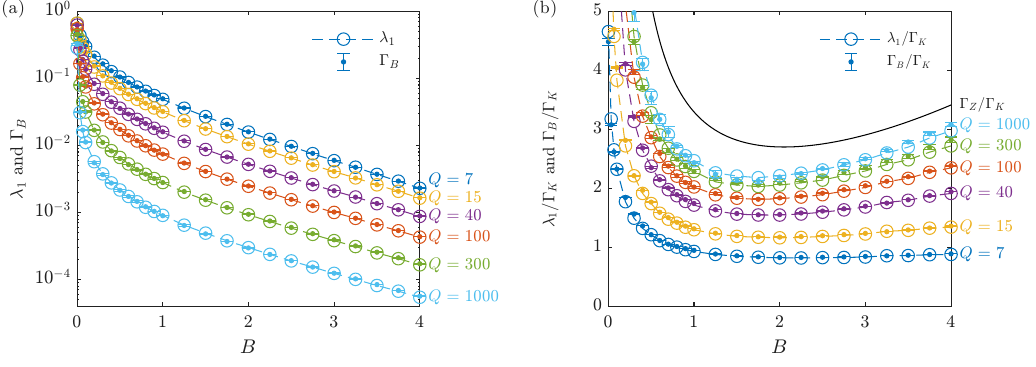} 
 \caption{(a) Eigenvalues $\lambda_1$ (circles and dashed line) of the differential operator $\mathcal{L}$ (Eq.~\ref{Eq:Kram:mod}), and long time rates $\Gamma_B$ (dots with tiny errors bars) of the escape process as obtained through numerical integration of the Langevin equation \eqref{Lan:eq}, both as a function of $B$ for different values of $Q$. (b) Same data, rescaled by the Kramers' rate expectation $\Gamma_K=\e^{-B}/Q$. For large $Q$, the threshold crossing rate tends towards the asymptotic value $\Gamma_Z$ (continuous line, Eq.~\ref{eq:GammaZwanzigA}).}
 \label{fig:lambda}
\end{figure*}

For any set of initial conditions $z_0=(x_0, v_0)$, and any point in the phase space $z=(x, v)$, the Kramers equation for the probability distribution function $P(z, t| z_0)$ reads~\cite{Redner_2001}:
\begin{eqnarray}
&& \partial_t P(z, t| z_0)= \mathcal L P(z, t| z_0), \\
&&\mathcal L \,\bullet = \left[-\partial_x v + \partial_v \left(x+\frac{v}{Q} + \frac 1 Q \partial_v\right)\right]\,\bullet.
 \label{Eq:Kram}
\end{eqnarray} 
We are interested in the absorbing boundary condition such that the Brownian particle starting at $z_0$ at $t=0$ is absorbed at $x=x_B$ for any value of $v$. Given that we take $x_B>0$ the particle will cross the boundary with positive velocity $v>0$. The absorbing condition can be included in the Kramers differential operator so that it reads: 
\begin{eqnarray}
&& \partial_t P(z, t| z_0)= \mathcal{L}_{x_B}P(z, t| z_0), \\
&& \mathcal{L}_{x_B}\,\bullet=\bigg[-\partial_x v + \partial_v \left(x+\frac{v}{Q} + \frac 1 Q \partial_v\right)\nonumber\\
&& \qquad \qquad \qquad \qquad \qquad \qquad -v \delta(x-x_B)\bigg]\, \bullet. \label{Eq:Kram:mod}
\end{eqnarray}
The sink term in Eq.~\ref{Eq:Kram:mod} accounts for the discontinuity in the probability current at $x_B$, while keeping the density equal to zero for $x>x_B$~\cite{Gardiner2004}. Similarly to the Kramers operator $\mathcal L$, the operator $\mathcal{L}_{x_B}$ is not self-adjoint, so one has to resort to the expansion of the time-dependent solution of Eq.~\ref{Eq:Kram:mod} in the biorthogonal set of eigenfunctions of the operator itself~\cite{Risken}. Given the set of right and left eigenfunctions $\psi_n(z)$ and $\phi_n(z)$, respectively, with $\mathcal{L} \psi_n(z)=-\lambda_n \psi_n(z)$ and $\mathcal{L}^\dagger \phi_n(z)=-\lambda_n \phi_n(z)$, the solution to Eq.~\ref{Eq:Kram:mod} reads
\begin{equation}
P(z, t| z_0)= \sum_n \e^{-\lambda_n t} \psi_n(z) \phi_n(z_0), 
\label{Pz:eq}
\end{equation}
where we have taken an explicit sign minus when introducing the eigenvalues for practical reasons, as will be detailed below.

Thus,  given the initial condition $z_0$, the   survival probability reads 
\begin{equation}
 S(t, z_0)= \sum_n \e^{-\lambda_n t } \phi_n(z_0)\int d z\, \psi_n(z), 
\label{eq:St}
\end{equation}
and the first passage time distribution across $x_B$ is then 
\begin{equation}
 f(t, z_0)=-\partial_tS(t, z_0).
\label{eq:ft}
\end{equation}
Given that the operator $\mathcal{L}_{x_B}$ describes the escape of the Brownian particles from volume $x<x_B$, the total probability inside such a volume is not conserved, and thus all eigenvalues $-\lambda_n$ will have negative real parts. Ordering the eigenvalues by increasing real part $\Re{\lambda_n}<\Re{\lambda_{n+1}}$, we conclude that the survival probability and the first passage time are dominated by the eigenvalue $-\lambda_1$ in the long time limit. Inspection of Eqs.~\eqref{eq:St}--\eqref{eq:ft} indicates that the long time exponential behavior of both the survival probability and the first passage time distribution is unaffected by the initial condition $z_0$, thus averaging over an arbitrary initial probability distribution 
for $z_0$ one obtains 
\begin{equation}
 f(t)\sim \lambda_1 \e^{-\lambda_1 t }, 
\label{eq:ft:lt}
\end{equation}
and thus the Kramers escape rate in the long time limit is just $\Gamma_B=\lambda_1(x_B)$ where we have now made explicit the dependence on the position of the absorbing boundary. 

\begin{figure*}[t]
 \centering
 \includegraphics{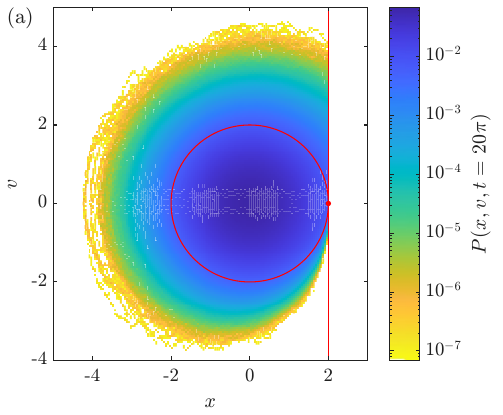}
 \includegraphics{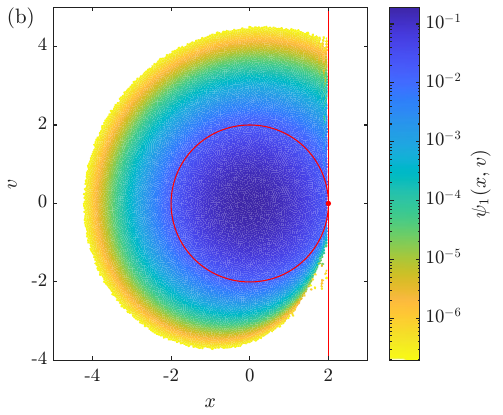}
 \caption{(a) Phase space density P(x, v, t) at large times ($t=20\pi\approx 4.5\trelax$), from direct numerical simulation of the Langevin equation for $x_B=2$ and $Q=7$. (b) Eigenfunction $\psi_1(x, v)$ corresponding to the slowest eigenvalue $\lambda_1$ of the differential operator $\mathcal{L}_{x_B}$ (Eq.~\ref{Eq:Kram:mod}), for $x_B=2$ and $Q=7$.}
 \label{fig:psi1}
\end{figure*}

The dominant eigenvalue $\lambda_1(x_B)$ can be obtained through standard numerical methods, namely the Finite Element Method that converts the continuous differential operator $\mathcal{L}_{x_B}$ into a finite-dimensional matrix problem. In Fig.~\ref{fig:lambda} we plot the results of such a numerical diagonalization of the operator $\mathcal{L}_{x_B}$, for different values of $B$, together with the escape rate in the long time limit as obtained from numerical simulation of the Langevin equation \eqref{Lan:eq}, finding an excellent agreement. In Fig.~\ref{fig:psi1}, we represent the eigenfunction $\psi_1(z)$ next to the phase space density computed from the Langevin numerical simulation for a large time, demonstrating again the excellent agreement: once all faster modes have decayed towards 0, the phase space is dominated by the exponential decay of the remaining density with the escape rate $\Gamma = \lambda_1$. Movies of the phase space evolution from numerical simulation of the Langevin equations are available as ancillary material~\cite{SuppMatMovies}, showing the transient toward this exponential decay state and its long time predominance.

\section{Large $Q$ limit} \label{Sec_large_Q}

In the limit of very large $Q$ (small friction), it is convenient to resort to polar coordinate, energy $E$ and phase $\theta$, with $x=\sqrt{2 E} \cos\theta$ and $v=\sqrt{2 E} \sin\theta$. The reason is that in the limit of very weak friction the energy is almost constant compared to the rapidly varying phase. In this limit one can thus average over the fast variable $\theta$ and obtain a Kramers-like equation for the energy alone~\cite{KRAMERS1940284, Zwanzig-2001, RevModPhys.62.251}
\begin{eqnarray}
&&\partial_t P(E, t| E_0)=\mathcal{L}_E P(E, t| E_0), \\
&&\mathcal{L}_E \, \bullet=-\partial_E [A(E)\, \bullet]+ \partial_E [D(E)\partial_E \, \bullet].
\end{eqnarray} 
where $D(E)$ and $A(E)$ are the energy diffusion coefficient and the drift term, respectively.
For the harmonic oscillator with quality factor $Q$ considered in the present paper, one finds $A(E)=-E/Q$ and $D(E)=E/Q$~\cite{Saadat_2023}.

The first passage time at an energy $E=B$ starting from a state with energy $E_0$ obeys the adjoint equation with boundary conditions~\cite{Zwanzig-2001}
\begin{subequations}\label{eq:adj}
\begin{align}
 \mathcal{L}^\dagger_{E_0}\tau(B,E_0)&=-1, \\
 \tau(B,E_0)|_{E_0=B}&=0, \label{eq:adjboundB}\\
 \partial_{E_0}\tau(B,E_0)|_{E_0=0}&=0 \label{eq:adjbound}
\end{align}
\end{subequations}
with $\mathcal{L}^\dagger_{E_0}$ the adjoint operator of $\mathcal{L}_E$ acting on the initial energy $E_0$
\begin{equation}
 \mathcal{L}^\dagger_{E_0}\bullet =\frac{1}{Q}\left[(1-E_0)\partial_{E_0} +E_0 \partial^2_{E_0}\right]\bullet \, ,
 \end{equation}
and where Eq.~\ref{eq:adjboundB} expresses instantaneous escape when the initial energy is equal to the value of the absorbing boundary $E_0=B$, and Eq.~\ref{eq:adjbound} ensures that no trajectory reaches a negative energy (reflecting boundary condition at $E_0=0$)~\cite{Saadat_2023}. Solution to Eq.~\ref{eq:adj} thus reads
\begin{eqnarray}
\tau_Z(B, E_0)&=& Q \int_{E_0}^B \d E\, \frac{\e^{E}-1}{E}.
\end{eqnarray} 

Finally, one can average over the initial probability distribution, which in the present paper corresponds to the equilibrium distribution for the states with $E<B$ , to obtain
\begin{align}
\tau_Z(B)&=\int_0^B \dd E_0\, \frac{\e^{-E_0}}{1-\e^{-B}}\, \tau_Z(B, E_0) \\
&=Q \int_0^B \dd E_0 \int_{E_0}^B \dd E\, \frac{\e^{-E_0}}{1-\e^{-B}} \frac{\e^E-1}{E}.
\end{align}

\begin{figure}[t]
	\centering
	\includegraphics{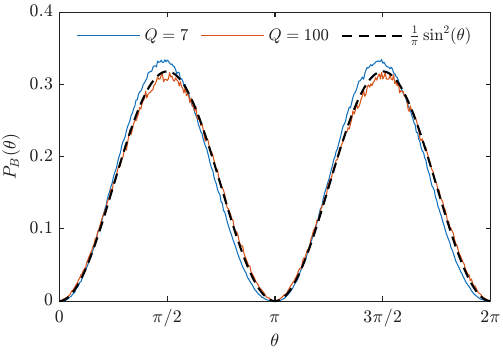}
	\caption{Distribution of phase $P_B(\theta)$ when exiting for the first time the circle defined by $E=B$, for $B=2$, $Q=7$ and $Q=100$. The black line corresponds to $P_B(\theta)= \frac{1}{\pi} \sin^2\theta$.}
	\label{Fig:histtheta}
\end{figure}

Once a thermal fluctuation has taken the system to an energy $B$, if damping is weak ($Q\gg1$) then in less than an oscillation period the position $x$ is very likely to cross the threshold $x_B$. To estimate the average time needed to reach $x_B$, let us compute the distribution of phases $\theta$ when reaching the energy level $B$. We start with the Kramers equation \eqref{Eq:Kram} which we rewrite in terms of the current $\mathbf{J}=(J_x, J_y)$ in the phase space:
\begin{align}
\partial_t P &= -\nabla\cdot \mathbf{J} = -\partial_xJ_x-\partial_v J_v, \\
J_x &= v P, \\
J_v &= -x P - \frac{1}{Q} vP - \frac{1}{Q} \partial_v P.
 \label{Eq:J}
\end{align}
The probability $P_B(\theta)$ to reach the energy threshold $E=B$ for a particular phase $\theta$ is computed by projecting the current $\mathbf{J}$ on the unit vector $\mathbf{n_\theta}=(\cos\theta, \sin\theta)$ normal to the circle of radius $\sqrt{2B}$ in the phase space:
\begin{align}
P_B(\theta) &\propto \mathbf{J} \cdot \mathbf{n_\theta}\\
&\propto - \frac{1}{Q} P \sqrt{2B} \sin^2\theta - \frac{1}{Q} \partial_v P \sin\theta
 \label{Eq:PB(theta)}
\end{align}
The first term in Eq.~\ref{Eq:PB(theta)} corresponds to the inward flux due to dissipation, and the second one to the outward flux due to the thermal noise forcing. For an harmonic oscillator in equilibrium, those two terms cancel since $\partial_v P = -vP$: the phase space is indeed stationary and the flux through a closed surface is zero. For the first passage time problem under scrutiny, quasi-stationariness of the flux through the circle of radius $\sqrt{2B}$ is a reasonable hypothesis, hence we expect $P_B(\theta) \sim \sin^2\theta$. This is supported in Fig.~\ref{Fig:histtheta} where $P_B(\theta)=\frac{1}{\pi} \sin^2\theta$ accurately describes the sampling of this pdf by numerical simulations for $Q=7$ and $Q=100$.

From $P_B(\theta)$, we easily compute the mean phase of the oscillator when crossing $E=B$ to be $\mean{\theta} = \pi$. Once a thermal fluctuation has taken the system to an energy $B$, it therefore takes a time $\pi$ in our reduced time units to reach the threshold at $x_B$, resulting in a first passage rate
\begin{equation} \label{eq:GammaZwanzigA}
 \Gamma_Z(x_B)=\frac{1}{\pi+\tau_Z(B)}.
\end{equation}

\begin{figure}[t]
	\centering
	\includegraphics{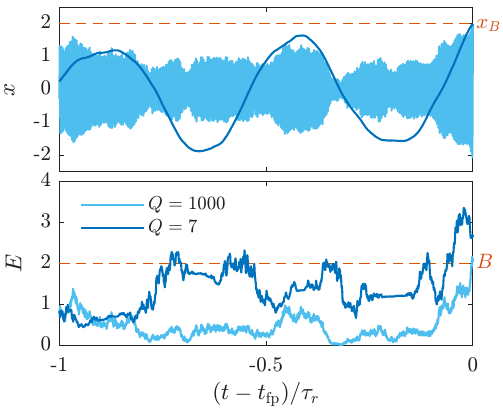}	
	\caption{Example of individual trajectories during one relaxation time $\trelax$ before reaching $x_B=2$ for $Q=7$ and $Q=1000$. Notice that at $Q=7$ the relaxation time corresponds to only two periods whereas it lasts about $300$ oscillation periods at $Q=1000$. The trajectories start with $E<B=\frac{1}{2}x_B^2$ and end at $t=\tfp$ when $x(\tfp)\ge x_B$. At small $Q$, $E>B$ can happen much earlier than $\tfp$, with $E$ crossing $B$ several time before $x_B$ is reached. At large $Q$, as soon as $E(t)>B$, $\tfp$ is reached in an average time around $\pi$ thanks to the small damping.}
	\label{Fig:E_v_t}
\end{figure}

As illustrated in Fig.~\ref{fig:lambda}(b), this line of reasoning only works in practice in the limit of very weak friction, i.e., large $Q$. For intermediate values of the quality factor where the energy is still fluctuating notably at the scale of one oscillation period, reaching $B$ does not ensure the crossing of the barrier at $x_B$: the energy crossing mean time $\tau_{Z}(B)$ underestimates significantly $\mean{\tfp(x_B)}$ (by a factor around 3 for $Q=7$ for example). The oscillator can cross the energy barrier $B$ several times before reaching the threshold at $x_B$, as one can conclude by inspecting in Fig.~\ref{Fig:E_v_t} a sample of stochastic trajectories obtained through the numerical integration of eq.~\eqref{Lan:eq}.

\section{Noise pattern generating an escape event}\label{Sec_noise_pattern}

In order to have a better understanding of the dynamics leading to $\tfp$ we consider all the trajectories starting at $x<x_B$. In Fig.~\ref{Fig:E_v_t} we plot the time evolution of $E(t)$ for $t\le \tfp$ for $Q=7$ and $Q=100$. As discussed in previous section, a first passage time based on energy diffusion (Eq.~\ref{eq:GammaZwanzigA}) is inaccurate at small $Q$. Indeed we clearly see that the energy can be larger than $B$ several times and long before $t=\tfp$ where $x\ge x_B$, due to the kinetic energy contribution. In contrast, we see that at large $Q$ $\tfp$ is very close to the first time $\tau_Z$ at which $E(\tau_Z)\ge B$, since due to the very small dissipation $x$ reach $x_B$ in half a period in average. The analysis of these trajectories gives a qualitative explanation of the different approaches at different $Q$ used in previous sections.

\begin{figure}[t]
	\centering
	\includegraphics{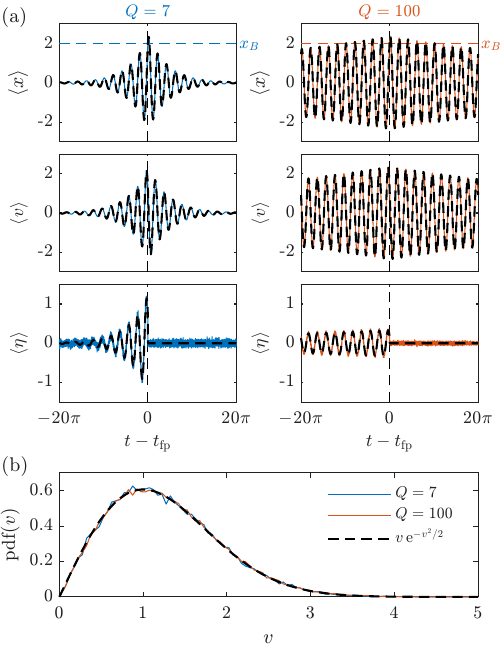} 	
	\caption{(a) Mean trajectories of the position $x$, velocity $v$ and noise $\eta$ (defined in Eq.~\ref{Lan:eq}) in an interval of time centered at $\tfp$ at $Q=7$ and $Q=100$. The model of Eq.~\ref{Eq:instanton} (with no free parameters) perfectly matches the numerical simulation data. Notice the very specific noise mean trajectory.
 (b) Distribution of velocities when crossing the threshold $x_B=2$, for two quality factors $Q=7$ and $Q=100$. The pdf is perfectly described by the theoretical expectation $\mathrm{pdf}(v)=v\e^{-v^2/2}$.
 }
	\label{Fig:noisepattern}
\end{figure}

In order to gain further insight into the kinetic energy contribution, we study the mean trajectories of $x$ and $v$ when crossing the threshold $x_B$. To this aim, we run numerical simulations of the dynamics of the harmonic oscillator (Eq.~\ref{Lan:eq}), without stopping them when $x$ crosses $x_B$ in order to follow the dynamics after the threshold as well. We then detect all crossings and average the trajectories on a short time interval centered in $\tfp$. From $\mean{x(t)}$, we can also compute the mean velocity $\mean{v(t)}$, acceleration, and finally the thermal noise term $\mean{\eta(t)}$ from Eq.~\ref{Lan:eq}. As pictured in Fig.~\ref{Fig:noisepattern}(a), these mean trajectories present a characteristic shape, with an exponential growth of a sinusoidal signal till $x$ reaches $x_B$, followed by exponential decay of the sinusoidal signal with no forcing.

We will now decode this specific pattern, starting with its final part ($t>\tfp$). Once the threshold has been reach, there are no correlation to be expected for the random forcing noise, so that its mean value averages to 0. The mean response of the linear system thus corresponds to the relaxation of an harmonic oscillator starting in $x(\tfp)=x_B$ with velocity $\mean{v(\tfp)}$. In Fig.~\ref{Fig:noisepattern}(b), we plot the distribution of velocities at $\tfp$, which is independent of the quality factor $Q$ and perfectly described by $P_{\tfp}(v)=v\, \e^{-v^2/2}$. Such expression is expected by noticing that $J_x = v\, P(x, v)$ is the projection of the phase space current $\mathbf{J}$ along the normal to the absorbing boundary $x=x_B$. The average velocity $\overline{v_B}$ on this boundary is therefore:
\begin{equation}
 \overline{v_B} \equiv \mean{v(\tfp)} = \int_0^{+\infty} v P_{\tfp}(v) \d v = \sqrt{\frac{\pi}{2}}
\end{equation}
The harmonic oscillator relaxation after $\tfp$ is then simply
\begin{equation}
 x(\Delta t)=\e^{-\frac{\Delta t}{2Q}}\left[x_B \cos\omega\Delta t+(\overline{v_B}-\frac{x_B}{2Q})\frac{\sin\omega\Delta t}{\omega}\right] \label{Eq:relax}
\end{equation}
where $\omega=\sqrt{1-1/4Q^2}$ and $\Delta t=t-\tfp$.

For $\Delta t>0$, the most probable relaxation path to leave the threshold is thus described by Eq.~\ref{Eq:relax}. For $\Delta t<0$, the most probable path to reach the threshold is the instanton matching the boundary condition $x(\tfp)=x_B$ and $v(\tfp)=\overline{v_B}$. In the harmonic oscillator case, this instanton is simply the time reversed trajectory of the relaxation~\cite{Bouchet-2014, Rosinberg_2016_EPL}, with initial conditions $x(\tfp)=x_B$ and $v(\tfp)=-\overline{v_B}$, which is described by Eq.~\ref{Eq:relax} simply switching the sign of $Q$. Eventually, the mean trajectory around $\tfp$ is described by this single formula:
\begin{align}
 x(\Delta t)=\exp\left(-\frac{|\Delta t|}{2Q}\right)\bigg[&x_B \cos\omega\Delta t\, + \label{Eq:instanton}\\
 &\left(\overline{v_B}-s(\Delta t)\frac{x_B}{2Q}\right)\frac{\sin\omega\Delta t}{\omega}\bigg] \ \ \ \ ~\nonumber 
\end{align}
with $s(t)=t/|t|$ the sign function. This theoretical expression is plotted as a black dashed line in Fig.~\ref{Fig:noisepattern}(a) and perfectly describes the mean trajectories $\mean{x(t)}$, $\mean{v(t)}$ and $\mean{\eta(t)}$ for any $Q$.

From this mean dynamics one deduces that in an underdamped system the escape events correspond to an increase of the noise around the resonance in a frequency band of the order of $1/Q$. The observation of this particular noise shape just before $\tfp$ might be used to define another method to estimate $\tfp$ based on the probability of having a given noise amplitude in a frequency band around the resonance.

\section{Initial transient} \label{Sec_short_FPT}

In the previous section, we focused of the long time behavior of the first passage distribution, leaving aside the transients that occur during the first oscillation periods. The companion letter~\cite{FPTuSHO_Letter} describes an approach to deal with times $t<2\pi$ (the first period), enough to handle intermediate dissipation regimes. For a quality factor $Q=7$ for example, the relaxation time $\trelax=2Q$ is only 2 periods, and even half of this value for energetic considerations~\cite{note-tauE}. All transients are therefore efficiently described by contributions $\PI$ to $\PIII$ in $P(\tfp)$ in Eqs.~\ref{Eq:pdftfp}--\ref{Eqs:PI-PII-PIII}. Besides $\PI$ and $\PII$, dealing with intrinsically short time contributions, an educated guess is used for contribution $\PIII$ described by Eq.~\ref{EqPIII}. Indeed, the escape rate $\Gamma(t)$ converges towards $\lambda_1(B)$ in the large time limit, and we know it is zero for $t<\pi$ since there is no contact between area III and the absorbing boundary $x=x_B$ in the phase space. For $\pi<t<2\pi$, the lowest energy of the contact line between areas III and I is $E^\dagger$. The ansatz we use is to consider that the rate of crossing the threshold $x_B$ for contribution at a short time $t<2\pi$ is can be approximated by the long time escape rate $\lambda_1(E^\dagger)$. Although this is clearly an approximation, it allows to have a closed form for the first passage distribution and is a very good match to the observed pdf at $Q=7$~\cite{FPTuSHO_Letter}.

\begin{figure}[t]
	\centering
	\includegraphics{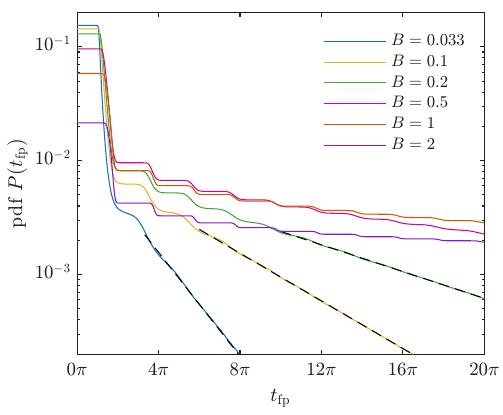} 	
	\caption{Distribution of first passage time $P(\tfp)$ from Langevin simulations for $Q=100$ and a few values of $B$ from $0.033$ to $2$ (see Appendix \ref{Appendix:Computepdf} for details). The Dirac contribution $\PI$ is not represented. A transient composed of several plateaux is visible before reaching the exponential decay regime (dashed black line plotted for the three smallest values of $B$).
 }
	\label{Fig:shortFPT}
\end{figure}

For larger $Q$, longer transients can occur before reaching the steady exponential decay state corresponding to the eigenstate $(\lambda_1, \psi_1)$ in the phase space. Such transients are clearly noticeable for $Q=100$ on the movies available as ancillary material~\cite{SuppMatMovies}, and in Fig.~\ref{Fig:shortFPT}. They manifest themselves as stair like decay, with a periodicity of one period ($2\pi$), and a vanishing step amplitude when $\tfp$ grows. This signature is a consequence of the eigenvalues $\lambda_{n>1}$, which have a higher value of their negative real part and decay faster than $\lambda_1$, but also a imaginary part multiple of the inverse of the period $1/2\pi$, inducing an oscillatory behavior in the first passage time distribution, see Eqs.~\eqref{Pz:eq}--\eqref{eq:ft}. The analysis on these transients could be performed in terms of the eigenvalues of the operator $\mathcal{L}_{x_B}$, but computing them goes beyond the scope of this article. As a general observation, we notice that the plateau pattern of the transients are more pronounced for large values of $B$ (see Fig.~\ref{Fig:shortFPT}), and typically last at most one relaxation time $\trelax$.

\section{Conclusions} \label{Sec_Conclu}
Our analysis of the FPTd in underdamped systems demonstrates the crucial role played by the quality factor in predicting the shape of the distribution. This analysis also shows the importance of studying separately the FPTs that are either short or long with respect to the relaxation time of the oscillator. Summarizing, we find that the FPTd is the sum of three parts. The first corresponds to zero FPT and the second to FPT less than the oscillation period. Both are independent of $Q$ and the presence of these two parts can be understood by studying the dynamics in the $(x, v)$ phase space in the Hamiltonian approximation. The third part depends on $Q$. At very large $Q$ the energy diffusion approach fully explains the exponential decays of the FPTd at large $\tfp$. Instead at moderate $Q$ the exponential decay rate of FPTd is determined by the slowest eigenvalue of the Kramers differential operator with absorbing boundaries. Our predictions fully agree with experimental~\cite{FPTuSHO_Letter} and numerical simulation data. In spite of the good agreement with observations, this is an average behavior because we have also shown that at intermediate $\tfp$ there is a reminiscence of the main oscillation period. Indeed the exponential decay is formed by several plateaux, each of duration $\sim2\pi$. Although the presence of these plateaux can be easily understood because of the oscillating dynamics, we have no method to determine them analytically. Finally we have observed that the dynamics of the system around the FPT is determined by specific noise patterns which drive the system along the mean escape trajectories. It will be interesting to understand how general this observation is in other systems and whether the presence of these specific pattern of the noise could be used as another method to compute the mean FPT in underdamped systems.

\acknowledgments
The data supporting this study will be available in an open public repository upon acceptance of the manuscript.\\

This work has been partially funded by project ANR-22-CE42-0022. We thank Satya Majumdar, Christopher Jarzynski and Corentin Herbert for enlightening scientific interactions. A.I. acknowledges support from the CNRS and the ENS de Lyon as an invited Researcher and invited Professor in the ENS de Lyon. 

\section*{Appendix}
\appendix

\begin{figure*}[t]
	\centering
	\includegraphics{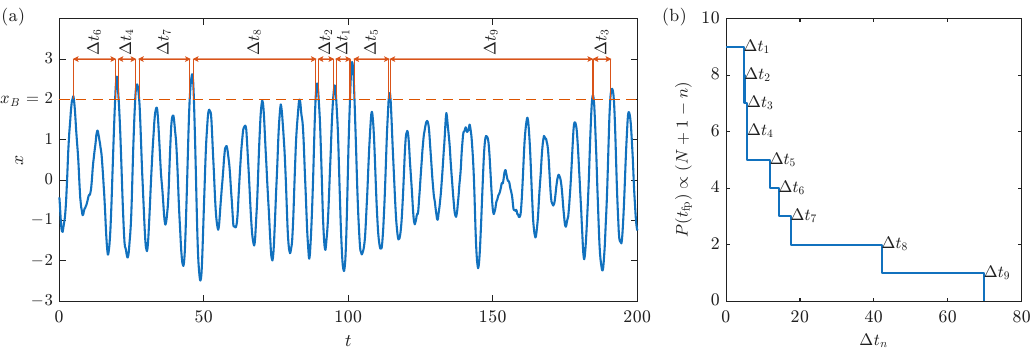}
	\caption{(a) Example of a time trace of $x$ from a numerical simulation of the Langevin equation \eqref{Lan:eq} with $Q=7$. We identify all time intervals $\Delta t_n$ where $x<x_B$ ($N=9$ intervals here with $x_B=2$), and sort them in ascending order. (b) The histogram of first passage times for this time trace can be reconstructed by plotting the decreasing linear range $N, N-1, \ldots , 2, 1$ as a function of $\Delta t_1, \Delta t_2, \ldots , \Delta t_{N-1}, \Delta t_N$.
	}
	\label{Fig:histtfp}
\end{figure*}

\section{Numerical method to compute $P(\tfp)$} \label{Appendix:Computepdf}

In this article, the data used to estimate the FPTd $P(\tfp)$ and plot Fig.~\ref{Fig:shortFPT}, compute $\Gamma_B$ in Fig.~\ref{fig:lambda}, or sample the phase space (Fig.~\ref{fig:psi1} and movies available as ancillary files~\cite{SuppMatMovies}) is generated with numerical simulation of the Langevin equation \eqref{Lan:eq} using a first order Euler integration scheme~\cite{Barros-2025} with a time step $\delta t = 2\pi/1000$. We draw initial conditions $(x_0,v_0)$ using the bi-variate Gaussian distribution corresponding to the equilibrium phase space of the harmonic oscillator, and run very long time trajectories in the harmonic potential, without interrupting the simulation when crossing the threshold. Thanks to ergodicity, such long trajectories sample the equilibrium phase space, and any point along them is a valid initial condition. We can {\it a posteriori} choose a threshold $x_B$ and look for the virtual first passage time $\tfp$ for any of these initial conditions. 

\begin{figure}[b]
    \centering
    \includegraphics{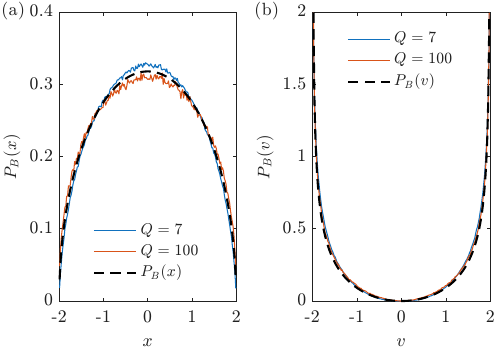}
	\caption{The pdfs of $x(\tau_Z)$ and $v(\tau_Z)$ are plotted in (a) and (b) respectively. The black bashed lines correspond to Eqs.~\ref{Eq:pdf_xtz} and \ref{Eq:pdf_vtz}.	}
	\label{Fig:pdf_theta_tz}
\end{figure}

The proportion of time above $x_B$ ($\tfp=0$) directly gives the Dirac contribution $\PI$. For $\tfp>0$, we proceed as follows. We first identify along the trajectory the residence times $\{\Delta t_n\}_{n=1\text{ to }N}$ below $x_B$, defined as the intervals of time where $x<x_B$. For one such interval $\Delta t_n$, any point along the trajectory is a valid initial condition, so that for any time $t^*<\Delta t_n$, there exist a sample trajectory crossing $x_B$ at $\tfp=t^*$. Let us now sort the $\{\Delta t_n\}$ is ascending order: $\Delta t_1<\Delta t_2<\cdots<\Delta t_N$. For any time $t^*<\Delta t_1$, there exist $N$ trajectories crossing $x_B$ at $\tfp=t^*$. For $\Delta t_1<t^*<\Delta t_2$, only $N-1$ trajectories allow for $\tfp=t^*$: the trajectories starting in $\Delta t_1$ are all too short and cross $x_B$ before $\Delta t_1$. For $\Delta t_{N-1}<t^*<\Delta t_N$, only $1$ trajectory allow for $\tfp=t^*$, and none above. The histogram of first passage time is thus obtained by plotting the decreasing linear range $N$ to $1$ as a function of $\{\Delta t_n\}_{n=1\text{ to }N}$, as illustrated in Fig.~\ref{Fig:histtfp}. We're only left with normalizing the histogram to estimate the pdf $P(\tfp)$, with a very high precision: we typically run simulation long enough to have $N\sim10^6$. The benefit of this approach is also that we can compute $P(\tfp)$ for any value of $x_B$ on the same dataset, without running the simulation again.

\section{Distributions $P(x)$ and $P(v)$ at $E=B$}

In section Sec.~\ref{Sec_large_Q} we have shown that the pdf of the angle $\theta$ at $t=\tau_Z$ is $P_B(\theta)={1\over \pi} \sin^2\theta$, where $\tau_Z$ is the time at which $E>B$ for the first time. In order to further check this result we discuss in this section the probability distributions $P_B(x)=P[x(\tau_Z)]\propto P_B(\theta)|\d\theta/\d x|$ and $P_B(v)=P[v(\tau_Z)]\propto P_B(\theta)|\d\theta/\d v|$. Taking into account that $x=\pm \sqrt{2B} \cos\theta$ and $v=\pm \sqrt{2B} \sin\theta$, we get 
\begin{eqnarray}
	P_B(x)&=& \frac{\sqrt{2B-x^2}}{\pi B},\label{Eq:pdf_xtz}\\
	P_B(v)&=& \frac{v^2}{\pi B\sqrt{2B-v^2}}. \label{Eq:pdf_vtz}	
\end{eqnarray}
Those distributions are in very good agreement with the numerical results as shown in Fig.~\ref{Fig:pdf_theta_tz}. 

\section{Link between the escape rate and the pdf of first passage time} \label{appendix-linkGammatfp}

Let us consider an initial population of $N_0$ systems, that cross the threshold at a rate $\Gamma(t)$. At any time $t$, the population $N(t)$ is decreasing as $N'(t)=-\Gamma(t) N(t)$, which can be integrated as 
\begin{equation}
	N(t)=N_0 \exp\left(-\int_0^t \Gamma(\tau)d\tau\right).
\end{equation}
By definition, the first passage time is the fraction of the population that overcomes the barrier at time $t$:
\begin{equation}
	P(\tfp)=-\frac{N'(\tfp)}{N_0}=\Gamma(\tfp)\exp\left(-\int_0^{\tfp} \Gamma(t)dt\right).
\end{equation}
Note that when $\Gamma=\Gamma_B$ is stationary (long time limit in the present article), then we simply have $P(\tfp) \propto \exp (-\Gamma_B \tfp)$. The prefactor $(1-\e^{-B})$ in Eq.~\ref{EqPIII} comes from the fraction of trajectories that are concerned by this escape rate, the rest overcoming the barrier in 
the first $2\pi$ period thanks to a large enough initial energy.

\bibliography{FPTuSHO}

\begin{thebibliography}{28}%
\makeatletter
\providecommand \@ifxundefined [1]{%
 \@ifx{#1\undefined}
}%
\providecommand \@ifnum [1]{%
 \ifnum #1\expandafter \@firstoftwo
 \else \expandafter \@secondoftwo
 \fi
}%
\providecommand \@ifx [1]{%
 \ifx #1\expandafter \@firstoftwo
 \else \expandafter \@secondoftwo
 \fi
}%
\providecommand \natexlab [1]{#1}%
\providecommand \enquote  [1]{``#1''}%
\providecommand \bibnamefont  [1]{#1}%
\providecommand \bibfnamefont [1]{#1}%
\providecommand \citenamefont [1]{#1}%
\providecommand \href@noop [0]{\@secondoftwo}%
\providecommand \href [0]{\begingroup \@sanitize@url \@href}%
\providecommand \@href[1]{\@@startlink{#1}\@@href}%
\providecommand \@@href[1]{\endgroup#1\@@endlink}%
\providecommand \@sanitize@url [0]{\catcode `\\12\catcode `\$12\catcode
  `\&12\catcode `\#12\catcode `\^12\catcode `\_12\catcode `\%12\relax}%
\providecommand \@@startlink[1]{}%
\providecommand \@@endlink[0]{}%
\providecommand \url  [0]{\begingroup\@sanitize@url \@url }%
\providecommand \@url [1]{\endgroup\@href {#1}{\urlprefix }}%
\providecommand \urlprefix  [0]{URL }%
\providecommand \Eprint [0]{\href }%
\providecommand \doibase [0]{https://doi.org/}%
\providecommand \selectlanguage [0]{\@gobble}%
\providecommand \bibinfo  [0]{\@secondoftwo}%
\providecommand \bibfield  [0]{\@secondoftwo}%
\providecommand \translation [1]{[#1]}%
\providecommand \BibitemOpen [0]{}%
\providecommand \bibitemStop [0]{}%
\providecommand \bibitemNoStop [0]{.\EOS\space}%
\providecommand \EOS [0]{\spacefactor3000\relax}%
\providecommand \BibitemShut  [1]{\csname bibitem#1\endcsname}%
\let\auto@bib@innerbib\@empty
\bibitem [{\citenamefont {Redner}(2001)}]{Redner_2001}%
  \BibitemOpen
  \bibfield  {author} {\bibinfo {author} {\bibfnamefont {S.}~\bibnamefont
  {Redner}},\ }\href@noop {} {\emph {\bibinfo {title} {A Guide to First-Passage
  Processes}}}\ (\bibinfo  {publisher} {Cambridge University Press},\ \bibinfo
  {year} {2001})\BibitemShut {NoStop}%
\bibitem [{\citenamefont {Chandrasekhar}(1943)}]{Chandrasek}%
  \BibitemOpen
  \bibfield  {author} {\bibinfo {author} {\bibfnamefont {S.}~\bibnamefont
  {Chandrasekhar}},\ }\bibfield  {title} {\bibinfo {title} {Stochastic problems
  in physics and astronomy},\ }\href {https://doi.org/10.1103/RevModPhys.15.1}
  {\bibfield  {journal} {\bibinfo  {journal} {Rev. Mod. Phys.}\ }\textbf
  {\bibinfo {volume} {15}},\ \bibinfo {pages} {1} (\bibinfo {year}
  {1943})}\BibitemShut {NoStop}%
\bibitem [{\citenamefont {Bray}\ and\ \citenamefont
  {Majumdar}(2013)}]{Bray_2013}%
  \BibitemOpen
  \bibfield  {author} {\bibinfo {author} {\bibfnamefont {A.~J.}\ \bibnamefont
  {Bray}}\ and\ \bibinfo {author} {\bibfnamefont {G.}~\bibnamefont {Majumdar},
  \bibfnamefont {S.~N.and~Schehr}},\ }\bibfield  {title} {\bibinfo {title}
  {Persistence and first-passage properties in nonequilibrium systems},\ }\href
  {https://doi.org/10.1080/00018732.2013.803819} {\bibfield  {journal}
  {\bibinfo  {journal} {Adv. Phys.}\ }\textbf {\bibinfo {volume} {62}},\
  \bibinfo {pages} {225} (\bibinfo {year} {2013})}\BibitemShut {NoStop}%
\bibitem [{\citenamefont {H\"anggi}\ \emph
  {et~al.}(1990{\natexlab{a}})\citenamefont {H\"anggi}, \citenamefont
  {Talkner},\ and\ \citenamefont {Borkovec}}]{Hanggi_1990}%
  \BibitemOpen
  \bibfield  {author} {\bibinfo {author} {\bibfnamefont {P.}~\bibnamefont
  {H\"anggi}}, \bibinfo {author} {\bibfnamefont {P.}~\bibnamefont {Talkner}},\
  and\ \bibinfo {author} {\bibfnamefont {M.}~\bibnamefont {Borkovec}},\
  }\bibfield  {title} {\bibinfo {title} {Reaction-rate theory: fifty years
  after kramers},\ }\href {https://doi.org/10.1103/RevModPhys.62.251}
  {\bibfield  {journal} {\bibinfo  {journal} {Rev. Mod. Phys.}\ }\textbf
  {\bibinfo {volume} {62}},\ \bibinfo {pages} {251} (\bibinfo {year}
  {1990}{\natexlab{a}})}\BibitemShut {NoStop}%
\bibitem [{\citenamefont {Majumdar}(2005)}]{Majumdar2005}%
  \BibitemOpen
  \bibfield  {author} {\bibinfo {author} {\bibfnamefont {S.}~\bibnamefont
  {Majumdar}},\ }\bibfield  {title} {\bibinfo {title} {Brownian functionals in
  physics and computer science},\ }\href
  {https://doi.org/10.1142/9789812772718_0006} {\bibfield  {journal} {\bibinfo
  {journal} {Curr. Sci.}\ }\textbf {\bibinfo {volume} {89}},\ \bibinfo {pages}
  {2076} (\bibinfo {year} {2005})}\BibitemShut {NoStop}%
\bibitem [{\citenamefont {Godec}\ and\ \citenamefont
  {Metzler}(2017)}]{Godec2017}%
  \BibitemOpen
  \bibfield  {author} {\bibinfo {author} {\bibfnamefont {A.}~\bibnamefont
  {Godec}}\ and\ \bibinfo {author} {\bibfnamefont {R.}~\bibnamefont
  {Metzler}},\ }\bibfield  {title} {\bibinfo {title} {First passage time
  statistics for two-channel diffusion},\ }\href
  {https://doi.org/10.1088/1751-8121/aa5204} {\bibfield  {journal} {\bibinfo
  {journal} {J. Phys. A: Math. Theor.}\ }\textbf {\bibinfo {volume} {50}},\
  \bibinfo {pages} {084001} (\bibinfo {year} {2017})}\BibitemShut {NoStop}%
\bibitem [{\citenamefont {Shin}\ and\ \citenamefont
  {Kolomeisky}(2019)}]{Shin2019}%
  \BibitemOpen
  \bibfield  {author} {\bibinfo {author} {\bibfnamefont {J.}~\bibnamefont
  {Shin}}\ and\ \bibinfo {author} {\bibfnamefont {A.~B.}\ \bibnamefont
  {Kolomeisky}},\ }\bibfield  {title} {\bibinfo {title} {Target search on dna
  by interacting molecules: First-passage approach},\ }\href
  {https://doi.org/10.1063/1.5123988} {\bibfield  {journal} {\bibinfo
  {journal} {J. Chem. Phys.}\ }\textbf {\bibinfo {volume} {151}},\ \bibinfo
  {pages} {125101} (\bibinfo {year} {2019})}\BibitemShut {NoStop}%
\bibitem [{\citenamefont {Majumdar}\ and\ \citenamefont
  {Ziff}(2008)}]{Majumdar_2008}%
  \BibitemOpen
  \bibfield  {author} {\bibinfo {author} {\bibfnamefont {S.~N.}\ \bibnamefont
  {Majumdar}}\ and\ \bibinfo {author} {\bibfnamefont {R.~M.}\ \bibnamefont
  {Ziff}},\ }\bibfield  {title} {\bibinfo {title} {Universal record statistics
  of random walks and l\'evy flights},\ }\href
  {https://doi.org/10.1103/PhysRevLett.101.050601} {\bibfield  {journal}
  {\bibinfo  {journal} {Phys. Rev. Lett.}\ }\textbf {\bibinfo {volume} {101}},\
  \bibinfo {pages} {050601} (\bibinfo {year} {2008})}\BibitemShut {NoStop}%
\bibitem [{\citenamefont {Evans}\ and\ \citenamefont
  {Majumdar}(2011)}]{Evans2011}%
  \BibitemOpen
  \bibfield  {author} {\bibinfo {author} {\bibfnamefont {M.~R.}\ \bibnamefont
  {Evans}}\ and\ \bibinfo {author} {\bibfnamefont {S.~N.}\ \bibnamefont
  {Majumdar}},\ }\bibfield  {title} {\bibinfo {title} {Diffusion with optimal
  resetting},\ }\href {https://doi.org/10.1088/1751-8113/44/43/435001}
  {\bibfield  {journal} {\bibinfo  {journal} {J. Phys. A: Math. Theor.}\
  }\textbf {\bibinfo {volume} {44}},\ \bibinfo {pages} {435001} (\bibinfo
  {year} {2011})}\BibitemShut {NoStop}%
\bibitem [{\citenamefont {Evans}\ \emph {et~al.}(2020)\citenamefont {Evans},
  \citenamefont {Majumdar},\ and\ \citenamefont {Schehr}}]{Evans2021}%
  \BibitemOpen
  \bibfield  {author} {\bibinfo {author} {\bibfnamefont {M.~R.}\ \bibnamefont
  {Evans}}, \bibinfo {author} {\bibfnamefont {S.~N.}\ \bibnamefont
  {Majumdar}},\ and\ \bibinfo {author} {\bibfnamefont {G.}~\bibnamefont
  {Schehr}},\ }\bibfield  {title} {\bibinfo {title} {Stochastic resetting and
  applications},\ }\href {https://doi.org/10.1088/1751-8121/ab7cfe} {\bibfield
  {journal} {\bibinfo  {journal} {J. Phys. A: Math. Theor.}\ }\textbf {\bibinfo
  {volume} {53}},\ \bibinfo {pages} {193001} (\bibinfo {year}
  {2020})}\BibitemShut {NoStop}%
\bibitem [{\citenamefont {Admon}\ \emph {et~al.}(2018)\citenamefont {Admon},
  \citenamefont {Rahav},\ and\ \citenamefont {Roichman}}]{Roichman_2018}%
  \BibitemOpen
  \bibfield  {author} {\bibinfo {author} {\bibfnamefont {T.}~\bibnamefont
  {Admon}}, \bibinfo {author} {\bibfnamefont {S.}~\bibnamefont {Rahav}},\ and\
  \bibinfo {author} {\bibfnamefont {Y.}~\bibnamefont {Roichman}},\ }\bibfield
  {title} {\bibinfo {title} {Experimental realization of an information machine
  with tunable temporal correlations},\ }\href
  {https://doi.org/10.1103/PhysRevLett.121.180601} {\bibfield  {journal}
  {\bibinfo  {journal} {Phys. Rev. Lett.}\ }\textbf {\bibinfo {volume} {121}},\
  \bibinfo {pages} {180601} (\bibinfo {year} {2018})}\BibitemShut {NoStop}%
\bibitem [{\citenamefont {Tal-Friedman}\ \emph {et~al.}(2020)\citenamefont
  {Tal-Friedman}, \citenamefont {Pal}, \citenamefont {Sekhon}, \citenamefont
  {Reuveni},\ and\ \citenamefont {Roichman}}]{Roichman_2020}%
  \BibitemOpen
  \bibfield  {author} {\bibinfo {author} {\bibfnamefont {O.}~\bibnamefont
  {Tal-Friedman}}, \bibinfo {author} {\bibfnamefont {A.}~\bibnamefont {Pal}},
  \bibinfo {author} {\bibfnamefont {A.}~\bibnamefont {Sekhon}}, \bibinfo
  {author} {\bibfnamefont {S.}~\bibnamefont {Reuveni}},\ and\ \bibinfo {author}
  {\bibfnamefont {Y.}~\bibnamefont {Roichman}},\ }\bibfield  {title} {\bibinfo
  {title} {Experimental realization of diffusion with stochastic resetting},\
  }\href {https://doi.org/10.1021/acs.jpclett.0c02122} {\bibfield  {journal}
  {\bibinfo  {journal} {J. Phys. Chem. Lett.}\ }\textbf {\bibinfo {volume}
  {11}},\ \bibinfo {pages} {7350} (\bibinfo {year} {2020})}\BibitemShut
  {NoStop}%
\bibitem [{\citenamefont {Vatash}\ and\ \citenamefont
  {Roichman}(2025)}]{Roichman_2025}%
  \BibitemOpen
  \bibfield  {author} {\bibinfo {author} {\bibfnamefont {R.}~\bibnamefont
  {Vatash}}\ and\ \bibinfo {author} {\bibfnamefont {Y.}~\bibnamefont
  {Roichman}},\ }\href@noop {} {\bibinfo {title} {Many-body colloidal dynamics
  under stochastic resetting: Competing effects of particle interactions on the
  steady state distribution}} (\bibinfo {year} {2025}),\ \Eprint
  {https://arxiv.org/abs/2504.10015} {arXiv:2504.10015 [cond-mat.soft]}
  \BibitemShut {NoStop}%
\bibitem [{\citenamefont {Besga}\ \emph {et~al.}(2020)\citenamefont {Besga},
  \citenamefont {Bovon}, \citenamefont {Petrosyan}, \citenamefont {Majumdar},\
  and\ \citenamefont {Ciliberto}}]{Besga_2020}%
  \BibitemOpen
  \bibfield  {author} {\bibinfo {author} {\bibfnamefont {B.}~\bibnamefont
  {Besga}}, \bibinfo {author} {\bibfnamefont {A.}~\bibnamefont {Bovon}},
  \bibinfo {author} {\bibfnamefont {A.}~\bibnamefont {Petrosyan}}, \bibinfo
  {author} {\bibfnamefont {S.~N.}\ \bibnamefont {Majumdar}},\ and\ \bibinfo
  {author} {\bibfnamefont {S.}~\bibnamefont {Ciliberto}},\ }\bibfield  {title}
  {\bibinfo {title} {Optimal mean first-passage time for a brownian searcher
  subjected to resetting: Experimental and theoretical results},\ }\href
  {https://doi.org/10.1103/PhysRevResearch.2.032029} {\bibfield  {journal}
  {\bibinfo  {journal} {Phys. Rev. Res.}\ }\textbf {\bibinfo {volume} {2}},\
  \bibinfo {pages} {032029} (\bibinfo {year} {2020})}\BibitemShut {NoStop}%
\bibitem [{\citenamefont {Besga}\ \emph {et~al.}(2021)\citenamefont {Besga},
  \citenamefont {Faisant}, \citenamefont {Petrosyan}, \citenamefont
  {Ciliberto},\ and\ \citenamefont {Majumdar}}]{Besga_2021}%
  \BibitemOpen
  \bibfield  {author} {\bibinfo {author} {\bibfnamefont {B.}~\bibnamefont
  {Besga}}, \bibinfo {author} {\bibfnamefont {F.}~\bibnamefont {Faisant}},
  \bibinfo {author} {\bibfnamefont {A.}~\bibnamefont {Petrosyan}}, \bibinfo
  {author} {\bibfnamefont {S.}~\bibnamefont {Ciliberto}},\ and\ \bibinfo
  {author} {\bibfnamefont {S.~N.}\ \bibnamefont {Majumdar}},\ }\bibfield
  {title} {\bibinfo {title} {Dynamical phase transition in the first-passage
  probability of a brownian motion},\ }\href
  {https://doi.org/10.1103/PhysRevE.104.L012102} {\bibfield  {journal}
  {\bibinfo  {journal} {Phys. Rev. E}\ }\textbf {\bibinfo {volume} {104}},\
  \bibinfo {pages} {L012102} (\bibinfo {year} {2021})}\BibitemShut {NoStop}%
\bibitem [{\citenamefont {Faisant}\ \emph {et~al.}(2021)\citenamefont
  {Faisant}, \citenamefont {Besga}, \citenamefont {Petrosyan}, \citenamefont
  {Ciliberto},\ and\ \citenamefont {Majumdar}}]{Faisant_2021}%
  \BibitemOpen
  \bibfield  {author} {\bibinfo {author} {\bibfnamefont {F.}~\bibnamefont
  {Faisant}}, \bibinfo {author} {\bibfnamefont {B.}~\bibnamefont {Besga}},
  \bibinfo {author} {\bibfnamefont {A.}~\bibnamefont {Petrosyan}}, \bibinfo
  {author} {\bibfnamefont {S.}~\bibnamefont {Ciliberto}},\ and\ \bibinfo
  {author} {\bibfnamefont {S.~N.}\ \bibnamefont {Majumdar}},\ }\bibfield
  {title} {\bibinfo {title} {Optimal mean first-passage time of a brownian
  searcher with resetting in one and two dimensions: experiments, theory and
  numerical tests},\ }\href {https://doi.org/10.1088/1742-5468/ac2cc7}
  {\bibfield  {journal} {\bibinfo  {journal} {J. Stat. Mech.}\ }\textbf
  {\bibinfo {volume} {2021}},\ \bibinfo {pages} {113203} (\bibinfo {year}
  {2021})}\BibitemShut {NoStop}%
\bibitem [{\citenamefont {Archambault}\ \emph
  {et~al.}(2026{\natexlab{a}})\citenamefont {Archambault}, \citenamefont
  {Crauste-Thibierge}, \citenamefont {Imparato}, \citenamefont {Ciliberto},\
  and\ \citenamefont {Bellon}}]{FPTuSHO_Letter}%
  \BibitemOpen
  \bibfield  {author} {\bibinfo {author} {\bibfnamefont {A.}~\bibnamefont
  {Archambault}}, \bibinfo {author} {\bibfnamefont {C.}~\bibnamefont
  {Crauste-Thibierge}}, \bibinfo {author} {\bibfnamefont {A.}~\bibnamefont
  {Imparato}}, \bibinfo {author} {\bibfnamefont {S.}~\bibnamefont
  {Ciliberto}},\ and\ \bibinfo {author} {\bibfnamefont {L.}~\bibnamefont
  {Bellon}},\ }\href {https://doi.org/10.48550/arXiv.2607.01404} {\bibinfo
  {title} {First passage time for an underdamped harmonic oscillator and
  application to the power of an information engine}} (\bibinfo {year}
  {2026}{\natexlab{a}}),\ \bibinfo {note} {companion letter, with a focus on
  short time behavior and an application to the power of an information
  engine},\ \Eprint {https://arxiv.org/abs/2607.01404} {arXiv:2607.01404
  [cond-mat.stat-mech]} \BibitemShut {NoStop}%
\bibitem [{\citenamefont {Archambault}\ \emph
  {et~al.}(2026{\natexlab{b}})\citenamefont {Archambault}, \citenamefont
  {Crauste-Thibierge}, \citenamefont {Imparato}, \citenamefont {Ciliberto},\
  and\ \citenamefont {Bellon}}]{SuppMatMovies}%
  \BibitemOpen
  \bibfield  {author} {\bibinfo {author} {\bibfnamefont {A.}~\bibnamefont
  {Archambault}}, \bibinfo {author} {\bibfnamefont {C.}~\bibnamefont
  {Crauste-Thibierge}}, \bibinfo {author} {\bibfnamefont {A.}~\bibnamefont
  {Imparato}}, \bibinfo {author} {\bibfnamefont {S.}~\bibnamefont
  {Ciliberto}},\ and\ \bibinfo {author} {\bibfnamefont {L.}~\bibnamefont
  {Bellon}},\ }\href@noop {} {\bibinfo {title} {{Ancillary movies obtained by
  direct numerical simulations of the Langevin Eq.~\ref{Lan:eq}, showing the
  phase space evolution for $B=1$ and $B=2$, with two examples of quality
  factors: $Q=7$ and $Q=100$}}},\ \bibinfo {howpublished}
  {\href{https://arxiv.org/src/2607.01405/anc}{arxiv.org/src/2607.01405/anc}}
  (\bibinfo {year} {2026}{\natexlab{b}})\BibitemShut {NoStop}%
\bibitem [{not()}]{note-tauE}%
  \BibitemOpen
  \href@noop {} {}\bibinfo {note} {For energy that is quadratic in position and
  speed, the relaxation time is half of $\trelax$: $\trelax^E =
  \trelax/2=Q$}\BibitemShut {NoStop}%
\bibitem [{\citenamefont {Gardiner}(2004)}]{Gardiner2004}%
  \BibitemOpen
  \bibfield  {author} {\bibinfo {author} {\bibfnamefont {C.~W.}\ \bibnamefont
  {Gardiner}},\ }\href@noop {} {\emph {\bibinfo {title} {Handbook of Stochastic
  Methods for Physics, Chemistry, and the Natural Sciences}}},\ \bibinfo
  {edition} {3rd}\ ed.,\ Springer Series in Synergetics\ (\bibinfo  {publisher}
  {Springer},\ \bibinfo {year} {2004})\BibitemShut {NoStop}%
\bibitem [{\citenamefont {Risken}(1996)}]{Risken}%
  \BibitemOpen
  \bibfield  {author} {\bibinfo {author} {\bibfnamefont {H.}~\bibnamefont
  {Risken}},\ }\href {https://books.google.dk/books?id=MG2V9vTgSgEC} {\emph
  {\bibinfo {title} {The Fokker-Planck Equation: Methods of Solution and
  Applications}}},\ Springer Series in Synergetics\ (\bibinfo  {publisher}
  {Springer Berlin Heidelberg},\ \bibinfo {year} {1996})\BibitemShut {NoStop}%
\bibitem [{\citenamefont {Kramers}(1940)}]{KRAMERS1940284}%
  \BibitemOpen
  \bibfield  {author} {\bibinfo {author} {\bibfnamefont {H.}~\bibnamefont
  {Kramers}},\ }\bibfield  {title} {\bibinfo {title} {Brownian motion in a
  field of force and the diffusion model of chemical reactions},\ }\href
  {https://doi.org/https://doi.org/10.1016/S0031-8914(40)90098-2} {\bibfield
  {journal} {\bibinfo  {journal} {Physica}\ }\textbf {\bibinfo {volume} {7}},\
  \bibinfo {pages} {284} (\bibinfo {year} {1940})}\BibitemShut {NoStop}%
\bibitem [{\citenamefont {Zwanzig}(2001)}]{Zwanzig-2001}%
  \BibitemOpen
  \bibfield  {author} {\bibinfo {author} {\bibfnamefont {R.}~\bibnamefont
  {Zwanzig}},\ }\href@noop {} {\emph {\bibinfo {title} {Nonequilibrium
  Statistical Mechanics}}}\ (\bibinfo  {publisher} {Oxford University Press},\
  \bibinfo {address} {New-York},\ \bibinfo {year} {2001})\BibitemShut {NoStop}%
\bibitem [{\citenamefont {H\"anggi}\ \emph
  {et~al.}(1990{\natexlab{b}})\citenamefont {H\"anggi}, \citenamefont
  {Talkner},\ and\ \citenamefont {Borkovec}}]{RevModPhys.62.251}%
  \BibitemOpen
  \bibfield  {author} {\bibinfo {author} {\bibfnamefont {P.}~\bibnamefont
  {H\"anggi}}, \bibinfo {author} {\bibfnamefont {P.}~\bibnamefont {Talkner}},\
  and\ \bibinfo {author} {\bibfnamefont {M.}~\bibnamefont {Borkovec}},\
  }\bibfield  {title} {\bibinfo {title} {Reaction-rate theory: fifty years
  after kramers},\ }\href {https://doi.org/10.1103/RevModPhys.62.251}
  {\bibfield  {journal} {\bibinfo  {journal} {Rev. Mod. Phys.}\ }\textbf
  {\bibinfo {volume} {62}},\ \bibinfo {pages} {251} (\bibinfo {year}
  {1990}{\natexlab{b}})}\BibitemShut {NoStop}%
\bibitem [{\citenamefont {Saadat}\ \emph {et~al.}(2023)\citenamefont {Saadat},
  \citenamefont {Latella},\ and\ \citenamefont {Ruffo}}]{Saadat_2023}%
  \BibitemOpen
  \bibfield  {author} {\bibinfo {author} {\bibfnamefont {E.}~\bibnamefont
  {Saadat}}, \bibinfo {author} {\bibfnamefont {I.}~\bibnamefont {Latella}},\
  and\ \bibinfo {author} {\bibfnamefont {S.}~\bibnamefont {Ruffo}},\ }\bibfield
   {title} {\bibinfo {title} {Lifetime of locally stable states near a phase
  transition in the thirring model},\ }\href
  {https://doi.org/10.1088/1742-5468/acecf9} {\bibfield  {journal} {\bibinfo
  {journal} {J. Stat. Mech.}\ }\textbf {\bibinfo {volume} {2023}},\ \bibinfo
  {pages} {083207} (\bibinfo {year} {2023})}\BibitemShut {NoStop}%
\bibitem [{\citenamefont {Bouchet}\ \emph {et~al.}(2014)\citenamefont
  {Bouchet}, \citenamefont {Nardini},\ and\ \citenamefont
  {Tangarifec}}]{Bouchet-2014}%
  \BibitemOpen
  \bibfield  {author} {\bibinfo {author} {\bibfnamefont {F.}~\bibnamefont
  {Bouchet}}, \bibinfo {author} {\bibfnamefont {C.}~\bibnamefont {Nardini}},\
  and\ \bibinfo {author} {\bibfnamefont {T.}~\bibnamefont {Tangarifec}},\
  }\bibfield  {title} {\bibinfo {title} {Non-equilibrium statistical mechanics
  of the stochastic {N}avier-{S}tokes equations and geostrophic turbulence},\
  }in\ \href {https://doi.org/10.31338/uw.9788323517399.pp.3-68} {\emph
  {\bibinfo {booktitle} {5th Warsaw School of Statistical Physics}}},\ \bibinfo
  {editor} {edited by\ \bibinfo {editor} {\bibfnamefont {B.}~\bibnamefont
  {Cichocki}}, \bibinfo {editor} {\bibfnamefont {M.}~\bibnamefont
  {Napi{\'o}rkowski}},\ and\ \bibinfo {editor} {\bibfnamefont {J.}~\bibnamefont
  {Piasecki}}}\ (\bibinfo  {publisher} {Warsaw University Press},\ \bibinfo
  {year} {2014})\ p.~\bibinfo {pages} {3}\BibitemShut {NoStop}%
\bibitem [{\citenamefont {Rosinberg}\ \emph {et~al.}(2016)\citenamefont
  {Rosinberg}, \citenamefont {Tarjus},\ and\ \citenamefont
  {Munakata}}]{Rosinberg_2016_EPL}%
  \BibitemOpen
  \bibfield  {author} {\bibinfo {author} {\bibfnamefont {M.~L.}\ \bibnamefont
  {Rosinberg}}, \bibinfo {author} {\bibfnamefont {G.}~\bibnamefont {Tarjus}},\
  and\ \bibinfo {author} {\bibfnamefont {T.}~\bibnamefont {Munakata}},\
  }\bibfield  {title} {\bibinfo {title} {Heat fluctuations for underdamped
  langevin dynamics},\ }\href {https://doi.org/10.1209/0295-5075/113/10007}
  {\bibfield  {journal} {\bibinfo  {journal} {{EPL} (Europhysics Letters)}\
  }\textbf {\bibinfo {volume} {113}},\ \bibinfo {pages} {10007} (\bibinfo
  {year} {2016})}\BibitemShut {NoStop}%
\bibitem [{\citenamefont {Barros}\ \emph {et~al.}(2025)\citenamefont {Barros},
  \citenamefont {Whitelam}, \citenamefont {Ciliberto},\ and\ \citenamefont
  {Bellon}}]{Barros-2025}%
  \BibitemOpen
  \bibfield  {author} {\bibinfo {author} {\bibfnamefont {N.}~\bibnamefont
  {Barros}}, \bibinfo {author} {\bibfnamefont {S.}~\bibnamefont {Whitelam}},
  \bibinfo {author} {\bibfnamefont {S.}~\bibnamefont {Ciliberto}},\ and\
  \bibinfo {author} {\bibfnamefont {L.}~\bibnamefont {Bellon}},\ }\bibfield
  {title} {\bibinfo {title} {Learning efficient erasure protocols for an
  underdamped memory},\ }\href {https://doi.org/10.1103/PhysRevE.111.044114}
  {\bibfield  {journal} {\bibinfo  {journal} {Phys. Rev. E}\ }\textbf {\bibinfo
  {volume} {111}},\ \bibinfo {pages} {044114} (\bibinfo {year}
  {2025})}\BibitemShut {NoStop}%
\end{thebibliography}%

\end{document}